\documentclass{article}

\usepackage{amssymb,amsmath,xcolor,graphicx,xspace,colortbl,ragged2e,rotating} %
\usepackage{amsmath}  
\usepackage{amssymb,amsmath,xcolor,graphicx,xspace,colortbl,ragged2e,rotating}  
\usepackage{boxedminipage}  
\usepackage{graphics}  
\usepackage{ragged2e}  
\usepackage{wrapfig}  
\usepackage{xcolor}  
\graphicspath{{How_Why_1_swp_graphics/}{How_Why_1_swp_tcache/}{How_Why_1_swp_gcache/}}
\DeclareGraphicsExtensions{.pdf,.eps,.ps,.png,.jpg,.jpeg}
\graphicspath{{How_and_Why_swp_graphics/}{How_and_Why_swp_tcache/}{How_and_Why_swp_gcache/}}
\DeclareGraphicsExtensions{.pdf,.eps,.ps,.png,.jpg,.jpeg}
\graphicspath{{How_and_Why_did_2_graphics/}{How_and_Why_did_2_tcache/}{How_and_Why_did_2_gcache/}}
\DeclareGraphicsExtensions{.pdf,.eps,.ps,.png,.jpg,.jpeg}
\graphicspath{{How_and_Why_did_graphics/}{How_and_Why_did_tcache/}{How_and_Why_did_gcache/}}
\DeclareGraphicsExtensions{.pdf,.eps,.ps,.png,.jpg,.jpeg}

\begin{document}
\title{How and Why Did Probability Theory Come About? }
\author{\normalsize by \\\relax Nozer D. Singpurwalla \\\relax The City University of Hong Kong, Hong Kong, \\\relax  and The George Washington University, Washington, D.C.  \\\relax Boya
Lai  \\\relax The City University of Hong Kong, Hong Kong}
\maketitle

\begin{abstract}{\normalsize This paper is a top down historical perspective on the several phases in
the development of probability from its prehistoric origins
to its modern day evolution, as one of the key methodologies in artificial intelligence, data
science, and machine learning.
\\\relax  It is written in honor of Barry Arnold's birthday for his many contributions to
statistical theory and methodology. Despite the fact that much
of Barry's work is technical, a descriptive document to mark his achievements should not be
viewed as being out of line. Barry's dissertation adviser
at Stanford (he received a Ph.D. in Statistics there) was a philosopher of Science who dug deep
in the foundations and roots of probability, and it is this breadth of perspective is what Barry has
inherent. The paper is based on lecture materials compiled by the first author from various
published sources, and over a long period of time. The material below gives a limited list of references, because the cast of characters is many, and their contributions
are a part of the historical heritage of those of us who are interested in probability, statistics,
and the many topics they have spawned.
} \\[20pt]
\end{abstract}\newpage

\section{Overview
}
The material here attempts to give a top down historical perspective on the several
phases in the evolution of probability, from its prehistoric
origins for imperial needs, to its current state as a branch of mathematics. As a branch of the
mathematical sciences, probability evolved in five stages,
not counting a period of stagnation between the second and the third stages, when doubts were cast
about its relevance as a mathematical discipline. Also
pointed out are paradoxes in probability, spawned by the absence of its precise definition, leading
to its last two phases, namely, an axiomatic and a subjective view of probability.

\section{The Prehistoric Phase
}
The rulers of ancient Egypt, Greece, and Rome, collected census data for taxes,
grain distribution, and other matters of administration; this
activity certainty had an impact on the origins of probability.

Next to come was the ``Doomsday'' list of William the Norman (1027
-- 1087), which was so exhaustive an economic survey that it reminded one of the final and the last
judgment by God in the Christian faith. Following this were the ``London Bills of Mortality'',
published since 1517, and notions such as the \textit{chance of death} in a given time
period, the \textit{chance of survival }to a certain age, and the like, originated, around
about 1535, almost a century before John Graunt's
celebrated mortality table. Another impetus to the origins of probability came from marine insurance
in the 1300's, and also during the renaissance, wherein
an emphasis was placed on observation and experiments in the natural sciences -- especially, on
errors of observation.

From a philosophical angle, the interrelations between chance and causality have
been on the philosopher's agenda since the ancient times. These too had an impact on the
origins of probability. In 1292, a treatise on the theory of the logical ideas of
\textit{Syadvada} (which is the basis of India's Jaina religion), lists
seven predications of which the fourth supplies a foundation for modern probability.

Another influential angle was the famous dictum
of Thomas Hobbs (1588 -- 1679), whose thesis was that no matter for how long we observe a
phenomenon, this is not sufficient grounds for its absolute and definitive knowledge.

To summarize, the prehistoric impact on probability came from: census, commerce,
renaissance, scientific observation, and philosophy.

\section{Was Probability Not Spawned by Gambling?
}
Apart from the discussion above, there is another belief that probability theory
owes its birth to gambling. To some, this is a questionable
issue. They claim that since gambling has been practiced since 5000 BC, it could not have taken 6000
years for it to influence probability. Their view is
that it was commerce that really influenced the development of probability.

Nonetheless, gambling has had an impact on probability,
and its earliest traces are in the literature, such as ``\textit{De Vetula}'' of
Richard de Fournival (1200 -1250), and Dante's ``\textit{Divine Comedy}''
(1307 -- 1321), wherein combinatorial arguments pertaining to outcomes of games of chance were
mentioned.

Paccioli (1445 -1514) published in 1487 ``\textit{Summa de Arithmetica, Geometria,
Proportioni et Proportionalita}'', which was an encyclopedia of the mathematical
knowledge of his period in Venice, and in the section labelled ``unusual problems'', he
discussed the question of the fair division of stakes when a match is stopped
in advance of an agreed termination of the game. Paccioli's solution embeds notions of
probability. This is also called the ``\textit{problem of points}'', and was a
trigger point of the famous Pascal Fermat correspondence.

Cardano (1501 -- 1576), and Tartagalia (1499 -- 1557) contributed
much to the connection between probability and gambling. Cardano developed probabilistic notions in
``The Book on Games of Chance'', written in 1526, as
``\textit{Liber de Ludo Aleae}''. In this book, Cardano enumerates
possibilities, permutations, deviations of frequencies from ``portion'', introduces
the notions of fair games and expectation, equally likely events, and uses the addition and
multiplication rules of probability for independent events.
He even came close to inventing the law of large numbers. However, Cardano was an ardent gambler who
restricted his writings, only to games of chance. All the same, as one can surmise,
he set the stage for much that was to follow.

Tartagalia (1499-1557) published in Venice in 1556, his treatise on ``Number and
Measure'' in which he related problems of probability to those of combinatorics, and
offered correct solutions to the problems posed by Paccioli, in particular, the
problem of the division of stakes, (or the problem of points). Following Cardano and Tartagalia, was
Galileo (1564 -- 1642), who posited that errors of measurement are inevitable; they are symmetric,
and clustered around a true value. He in fact revealed many of the characteristics of the
normal probability distribution.

The above developments, perhaps mark the end of the phase of the earliest writings
on probability, subsequent to its prehistoric phase.

\section{ Development of Probability as a Science
}
This phase can be categorized into five stages, and includes a phase called ``the
period of stagnation'', between the second and the third stage, when
concerns were raised about probability as a branch of mathematics. Also included is a phase labeled
``paradoxes in probability'', which can be seen as
the doorstep to the development of the last two stages in the evolution of probability as a
mathematical discipline.

Within the five stages alluded to above, are also some milestones in the evolution
of statistics, which evolved as a way to reason with numbers.

\subsection{\underline {Stage I}. Of the Development of Probability as a
Science
}
Up until the middle of the 17th century, there were no general methods for solving
probabilistic problems. Specific problems had been solved,
and a substantial amount of knowledge was accumulated. The term probability (nor its disposition as
a number) was not a part of the lexicon in the solution of such problems.

In the middle of the 17th century some prominent mathematicians like Pascal,
Fermat, and Huygens became involved in the development of probability, even without mentioning the
term. These individuals were familiar with Cardano's addition and the multiplication rules,
the notion of independence, and put to practice the notion of expectation using combinatorics. They
developed new methods for solving problems, determined
the realm of problems to which this new science is applicable, and in so doing were on the verge of
transforming probability to a bona fide science.

Two very important and key individuals need to be mentioned in Stage I. They were:
Chevalier de M{\'e}re, and Christian Huygens. They brought
probability into a new stage as a science. Chevalier de M{\'e}re (1607 -- 1684) was a philosopher and a
man of letters; he wrote to Pascal about the division of stakes [considered by Paccioli but per Shafer (2019) solved in the 1400's by two Italian abacus masters, and his
(de M{\'e}re's) solution to it. With Pascal and Fermat, he had authored in 1662 ``\textit{Ars
Cognitandi}'' (Art of Thinking) as a part of the Arnold
- Nicole (who were abbots at the Port Royal Monastry), a book on ``\textit{Port
Royale}\textit{Logic}''. de M{\'e}re's letter to Pascal triggered a correspondence between Pascal
and Fermat in 1654, and thus originated the founding document on mathematical probability.

Even though many mathematicians of that period devoted much
attention to the solution of games of chance, actual gambling was condemned. Thus, the myth that
Chevalier de Mere was a fervent gambler. Rather, he was
a man of letters who viewed probability only as a ``useless curiosity''. By contrast,
Cardano, who was an ardent gambler, used mathematics for gambling,
but in 1526 did not quite hit upon the notion of probability as a number.

Christian Huygens (1629 -- 1695), a Dutchman from Holland,
visited Paris in 1655 to receive a doctorate in law. He was impressed with the problems on gambling
of Pascal and Fermat, and undertook further work on
it. He was told of the solutions but not the methods (which were published posthumously, because
both Pascal and Fermat posed problems to each other but
hid their methods of solution). The correspondence between Pascal and Fermat was published only in
1679.

Huygens returned to Holland and begun work on solving the problems posed by Pascal
and Fermat). Huygens solutions, independent of the methods of Pascal and Fermat, but identical to
those of Pascal and Fermat, were published in his book (written in Latin) called
``\textit{About}\textit{ Dice Games}''. This book appeared in 1657
wherein Huygens says ``... we are dealing not only with games but rather with a foundation of a new
theory, both deep and interesting.'' His reasons for writing this book was the
absence of methods used by Pascal and Fermat.

This book is viewed as the first published treatise on
mathematical probability. Huygens book can be viewed as being the first format document on the
introduction of mathematical probability, until Bernoulli's famous ``\textit{Ars
Conjectandi}'' (a possible imitation of the Pascal - Fermat- de Mere's, Ars
Cognitandi). Huygen's book was also the first to introduce and to apply
the notion of expectation in commercial and industrial problems. Huygen's terminology was
commercial.

Subsequent to the above, more and more works on probability began to appear,most
notable being the birth to a new discipline, now called ``\textit{Data
Science}''. In 1662, at about the same
time as Huygen's book, John Graunt, an Englishman, published a tiny book devoted to problems of
vital statistics. Huygens was asked to comment on this landmark
book, which he did favorably. Indeed, in 1669, using Graunt's work Huygens constructed a
mortality curve, and initiated the application of probability
to demography, and to annuities. In 1690, another Englishman, by the name of William Petty published
his treaties on ``\textit{Political Arithmetic}'', which
was about a method of reasoning on matters of government, via the use of numbers. This can now be
seen as a founding document on Government Statistics.

Preceding Petty's treatise, was work on actuarial mathematics and the worth
of annuities, due to deWitts in 1671, followed by that
of Edmund Haley in 1693, who published the very first mortality table based on data from Breslau.
Between (1791 - 1799), a Scotsman named John Sinclair
published 21 volumes of his Statistical Account of Scotland, and introduced the word
``\textit{Statistics}'' to replace Petty's political arithmetic.
Up until 1796 the word ``statistics'' was used in Germany to describe the political
strength, happiness, and the improvement of a country, as a measure
of its well-being. Statistics was an artificial word, with no evidential meaning, that is now used
for anything having to do with data. Sinclair used it
to garner attention over Petty's political arithmetic, which did not seemed to have gained
traction [cf. von Collani (2014).

To summarize, Huygens recognized the role of probability as a science, wrote the first book on it, applied the notion of
expectation to commerce and industry, and used probability
for assessing demography and insurance. Huygens' Book played an important role in the history
of probability. Jacob Bernoulli, who introduced the term ``probability'',
based on the Latin ``\textit{probabilitas}'', was greatly influenced by
Huygen's book. Bernoulli's work established the foundations of mathematical probability.

Bernoulli's word, probability is based on the term
``\textit{probabilitas}'', which was a moral system of the
Catholic Church. Probabilitas was formally introduced in 1577 by the Spanish Dominican, Bartholome
de Medina, and was mainly applied by Jesuit priests.
Bernoulli's aim in writing Ars Conjectandi was to introduce a new branch of science, that he
called Stochastics, or the science of prediction.

To Bernoulli, a relevant feature of ``stochastics'' was an event's
\textbf{\textit{readiness to occur}, }and ''probability''\textbf{, }the
\textbf{\textit{degree of certainty }}of its
occurrence, see von Colani (2014). Thus, to Bernoulli, stochastics was the art of measuring probability as exactly as is
possible.

However, Bernoulli acknowledged that the determination of the true
value of probability is impossible, and labeled as ``mad'' any attempt at doing so. This
motivated him to develop his law of large numbers, as an empirical
method to determine a lower and an upper limit for an unknown probability. Note that
Bernoulli's notion of probability was devoid of any mathematical basis.

Given below is a graphic of the evolution of mathematical probability, up until
the beginning of Stage II which established it as a mathematical science.

\begin{figure}[!ht]
  \centering
  \includegraphics[scale=0.8]{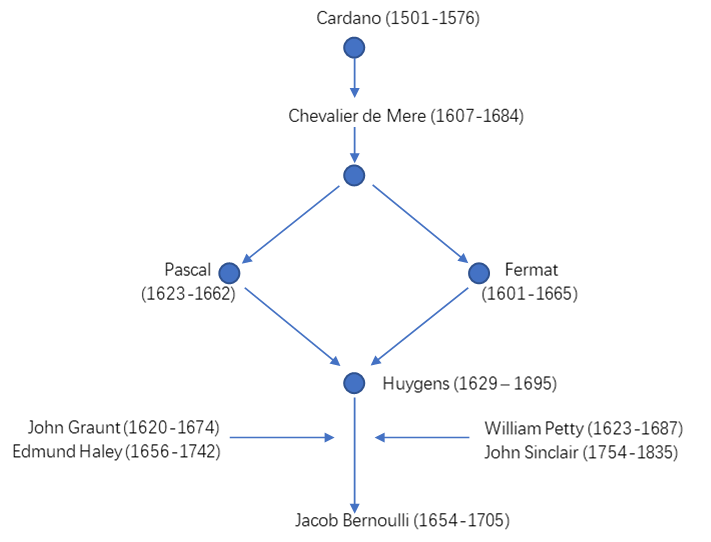}
  \caption{Evolution of Mathematical Probability}
\end{figure}

\subsection{\underline {Stage II}. Bernoulli Makes Probability a Bona Fide
Mathematical Science
}
James (Jacques) Bernoulli (1654 -1705) in his 1713 ``Ars Conjectandi''
proved the first limit theorem, and in so doing, raised the status of
probability to that of a formal mathematical science. This book was published posthumously by his
nephew Nichols (1) Bernoulli [who also applied probability
to matters of jurisprudence, like the credibility of a witness].

The contribution of Bernoulli to make probability a bona fide mathematical
science is that he interpreted propositions in Huygen's book, showed inapplicability of the
addition law to non-disjoint events, gave the binomial formula,
and used Leibnitz's combinatorics for solving probability problems.

He proved the weak law of large numbers as a way to bound a ``true''
probability, and interpreted probability as the degree of certainty of an event's occurrence.
Bernoulli was a metaphysical determinist; i.e. if we know
the position of a dice, its speed, its distance from the board, etc. we can exactly predict its
outcome. Thus, to Bernoulli, probability, or chance, the
terms he used interchangeably, depends on our state of knowledge, and is thus personal to the
individual specifying it.

However, since all knowledge is not possible,
we assume a statistical regularity in a large number of trials, say
$n$
, and conclude that for the tossing of coins, the deviation of m/n from p, as n
$ \rightarrow \infty $
, is small with a large probability;
$m$
is the total number of heads in the
$n$
trials. Bernoulli also touched upon the philosophical problems connected with
probability, and asserted probability should also be applied to situations outside games of chance.

Besides Bernoulli there were others who worked on probability during the beginning
of the 18-th century. We name a few.

Pierre de Montmort (1678 -- 1719), a mathematician, who was chosen by Leibniz on
the commission to inquire about the priority of inventing differential and integral calculus between
him and Newton; de Montmort favored Newton; see Maistrov (1974). His basic work on probability
entitled ``\textit{Essai d'Analyse sur les Jeux de Hazard}'' was published
in 1708 (5 years prior to Bernoulli's posthumously published work).
It was in a letter to de Montmort that Nicholas (1) Bernoulli posed the St. Petersburg paradox.
Montmort's main effort was the applications of probability to human behavior.

Abraham de Moivre's (1667 -- 1754) principal work in probability was in
``\textit{The Doctrine of Chances}'', 1718. Here,
without addressing the matter of what probability is, de Moiver discusses topics connected to
Bernoulli's theorem, and the problem of the duration of the play
(first proposed by Huygens). de Moivre investigated the probabilities of various deviations between
m/n and p, for p=1/2. Laplace extended these to p
$ \in $
(0,1), and thus the de Moivre-Laplace theorem is the second limit theorem in mathematical
probability.

Thomas Bayes (1792 -- 1761), speculated as being tutored by de Moivre, published
his famous essay posthumously in 1763; it was entitled ``\textit{Thomas
Bayes}'\textit{s Essay Towards }Solving \textit{a Problem in the Doctrine
of Chances}''; it addressed the following question: what is the chance that p
$ \in $
(a,b) given x and n? Bayes offered a solution to this problem using solely the calculus
of probability. In so doing, he introduced the notion of what is referred to as
``\textbf{\textit{probabilistic }\textit{induction}}''.

To obtain his solution, Bayes used what
is now called Bayes formula (which is really an alternative form of the well-known, by then,
multiplication rule), interpreted conditional probability and
its subtleties, and assumed a uniform distribution on p (via eliciting priors on the observables --
i.e. the predictive distribution). It was Laplace who
coined the term ``Bayes Theorem'' and set in notion this terminology -- Bayes did not
invent Bayes Theorem.

Daniel Bernoulli (1700 -- 1782) introduced the idea of probability curves, and
applied differential calculus to problems of probability theory, and in so doing simplified many
of the cumbersome combinatoric formulas used before. However, his most important contribution is the
introduction of the notion of ``\textit{utility}'' or ``\textit{moral
}expectation'', and its use in solving the St. Petersburg paradox, posed by
Nicholas (1) Bernoulli.

Condorcet [Jean Antoine de Caritat,
Marquis de Condorcet] (1743 -- 1794) was a well-known sociologist and economist during the period of
the French Revolution. His main contribution is his
introduction of the notion of ``\textit{probabilite' propre}'', which is a
subjective, or personal, probability. His ideas were rejected as being
beyond the scope of mathematical probability theory.

After Bernoulli, one of the great minds that came to wrestle with probability
was Pierre Simon. de Laplace. His main technical contribution is the de Moivre-Laplace central limit
theorem, for Bernoulli trials. His contribution to
larger issues is extending the realm of applicability of probability to social phenomena, and his
reinforcement of Condorcets notion of subjective probability.
That is ``probability is relative in part to ignorance, and our knowledge''. If a coin is
asymmetrical, but we do not know which side, then its probability
of head is 1/2. Laplace also played a role in developing statistics.

Stage II of the development of probability ends with Gauss (1777
-- 1855), who derived the normal law for the distribution of errors. [This was also done by Robert
Adrian (1755 -- 1843) an obscure American mathematician].

Poisson (1781 -1840) did much work on technical and practical aspects of
probability. He subscribed to the subjectivie view of
probability, and like Laplace felt that probability can also be applied to jurisprudence.
Poisson's main contributions is a generalization of Bernoulli's
theorem when the probability of an event changes from trial to trial, so that if
$\tilde{p}$
is the arithmetic mean of these probabilities, then
\begin{equation*}\lim_{n \rightarrow \infty }P (\vert \frac{m}{n} -\tilde{p}\vert  <\epsilon ) =1\text{,}
\end{equation*}
and his proof that as
$p_{n} \rightarrow 0$
, then as
$n \rightarrow \infty $
,
$P (m/  n) =\frac{e^{ -n}}{m !} e^{ -\lambda }$
, where
$\lambda  =n p_{n}$
, the famous Poisson formula; recall that m is the number of events in n Bernoulli
trials.

\subsection{ The Period of Stagnation
}
The period (1860-1900) is also viewed as one of stagnation in the development of
probability. Many felt that its application to social problems
was a compromise in the mathematical sciences.

The areas of application being not clearly defined there was much controversy
about the subject. There was much criticism of the early developers, like Pascal, Bernoulli,
Laplace, and Poisson for their subjectivist inklings via metaphysical
determinism. The period of stnagation terminated with the emergence of the now famous Russian School
of probability.

\subsection{\underline {Stage III}. Creation of The Russian School
}
The originators of the Russian School in probability were Ostrogradski
(1801-1862), and Bunyakowski (1804-1889). Ostrogradski, influenced by
Laplace, was a proponent of the principle of insufficient reason, and applied the theory of
probability to moral problems. He too subscribed to the notion
that probability is a measure of our ignorance, and is thus subjective. Bunyakowski wrote the first
Russian book in probability, and introduced the needed
terminology; he too was a determinist in the spirit of Bernoulli and Laplace.

Chebyshev (1821-1894), influenced by Ostrogradski and Bunyakowski,
is credited with the creation of the Russian school of probability. His students Markov, Voroni,
Lyapunov, and Steklov, pushed frontiers of the subject
to the modern era. In effect, Chebyshev and his followers, broke the period of stagnation and
impasse in probability, as a mathematical science. Chebyshev
defined the subject matter of probability theory as the mathematical science of constructing
probabilities of an event based on probabilities of other events.
He did not discuss how initial probabilities are to be obtained. Chebyshev introduced mathematical
rigor in the theorems, and obtained exact estimates or
inequalities of derivations from limiting laws which arise when the number of trials is large but
finite.

Philosophically, Chebychev and his followers were materialists through the natural
sciences, mechanics and, mathematics. They were guided by the opinion that only those investigations
initiated by applications are of value, and only theories which arise from a consideration of
particular cases are useful. The materialist philosophy is
founded on the belief that nothing exists but matter itself and its manifestations.

Markov (1856-1922) was Chebychev's closest disciples
and his most colorful spokesperson. He transformed probability, with clarity and rigor, to one of
the most perfect field in mathematics. His noteworthy
works are on the limit theorems for sums of independent and dependent random variables using the
method of moments. Markov introduced the famous chain named
after him, for analyzing Pushkin's poem ``\textit{Eugene Onegnin}''.

Lyapanov (1857-1918) improvised on the proofs of Markov's theorems using
characteristic functions; the central limit theorem is named after him. Lindberg and Feller later
improved on Lyapunov's theorems.

\subsection{ Probability in Physics
}
The evolution of probability as a mathematical science would not complete without
a mention of its impact in physics, one of the most basic of
all the sciences. In 1827, Robert Brown, an English botanist, detected the movement of minute
suspended particles in an unpredictable manner. This movement
is due to random bombardments of chaotically moving molecules in suspension. Using probabilistic
arguments, Albert Einstein in 1905 was able to develop
a sound theory for such motions. It was observed that every sufficiently small grain suspended in a
fluid constantly moves in an unpredictable manner.

If before the second half of the 19-th century, the basic areas of application of
probability were in the processing of observations, the second
half was in physics. This was prompted by the work of Ludwig Boltzman (1844-1906), an Austrian, and
Josiah Willard Gibbs (1839-1903), an American.

Boltzman is credited with the initiation of statistical physics, and the
probabilistic interpretation of entropy. His work paved the way for quantum
theory. Boltzman was preceded by Maxwell who thought of molecules as elastic solids, whose behavior
can be studied through the methods of probability.

In 1902 Gibbs, who was occupied with problems of mechanics, published his famous
book ``\textit{Basic Principles of Statistical Mechanics'}'. This book
was an influential development for the enhancement of probabilistic notions in physics.

\subsection{ Paradoxes in Probability
}
Towards the beginning of the 20-th century, great inroads were made in probability
as a mathematical discipline by Chebychev, Markov, and Lyapunov,
and into its inroads in physics by Maxwell, Boltman, and Gibbs. However, mathematicians were
repeatedly pointing out concerns regarding the need for a precise meaning of probability.

Indeed Bertrand (of Bertrand's Paradox, of which is Borel's Paradox, and
the three envelope problem are examples), and Henri Poincare, via their paradoxes tried to emphasize
the inaccuracies and vaguenesses in the basic notions of interpreting probability.

Emil Borel (1871-1956) and Henri Poincare (1854-1912), both prominent French
mathematicians, were determinists whose notion of probability, was
that it is a reflection of our ignorance. Both wrote two highly influential books on the subject,
and called for a rigorous definition of the meaning of
probability. These can be seen as paving the path towards, Stage IV and V, on the axiomatic and the
subjective development of probability.

\subsection{\underline {Stage IV}: The Axiomatic Development
}
The axiomatic method in science, particularly, the mathematical sciences, makes it
possible to apply any theory to many areas. For example, Lobachevskii
(1829) suggested the possibility of constructing geometry based on a system of axioms, different
from those of Euclid, whereas Hilbert, Peano, and Kagan,
investigated such a possibility for geometry in the early part of the 20-th century; Hilbert and
Peano also did this for arithmetic.

With probability, Laplace's classical definition using equiprobable events
was a tautology, because: equiprobable
$ \Leftrightarrow $
equal probability. Also, the subjective interpretation of probability had, at the early
part of the 20-th century, serious flaws having to do with a linear utility for money and state
dependence. As a consequence, the need for axiomatization was becoming more and more pressing.

In 1917, S.N. Bernstein (1880-1968) published a paper hinting the axiomatization
of probability. This marked a new stage in its development. Bernstein's axiomatization was
based on the notion of qualitative comparisons of events in which larger and smaller probabilities
serve as a foundation. Bernstein's ideas were further developed by Glivenko and more recently,
by Koopman (1940). Bernstein's notion of probability was
also materialistic, and was for applications to the natural sciences.

Richard von Mises (1883-1953) was a strong critic of both the equiprobable
and the subjective theory of interpreting probability. His main contribution is the frequency
approach; i.e. probability is relevant only to mass phenomena.
Approaches alternate to von Mises, were due to Keynes, followed by Harold Jeffreys, who viewed
probability as a degree of likelihood, wherein every proposition
has a certain definite probability. It is said that later on, Keynes recanted this position.

Simultaneous with attempts to lay the
foundations of probability were rapid new developments in the mathematical sciences, vis a vis the
works of Khinchin, Borel, Cantelli, Hardy, Littlewood,
and Hausdorff. These trends facilitated Kolmogorov to construct his axiomatization of probability
and lay the foundation for a decisive stage in its development.
In particular, Bernoulli's result on the weak law and Borel's on the strong law, led
Kolmogorov to notice the connection between probability and measure,
and thus began his work on axiomatization, resulting in the publication of his famous book, in
1933.

Kolmogorov's aim was not to clarify
the meaning of probability, but to establish a branch of mathematics in exactly the same way as
algebra and geometry. To Kolmogorov, the concept of a theory
of probability is a system of sets which satisfy certain conditions. He thus introduced the term
probability in the above context, detached from any real world meaning.

Not all applied scenarios satisfy Kolmogorov's set up and architecture.
Consequently, there are alternatives to probability
like Zadeh's Possibility Theory for fuzzy sets, and the Dempster-Shafer Belief Function
Theory.

\subsection{\underline {Stage V}: Personal Probability
}
Approaches at interpreting probability, alternate to the ``classical''
one of La Place, the ``frequency'' one of von Mises, the ``logical'' one
of Keynes, as well as the axiomatics of Kolmogorov (that technically speaking are free of
interpretation) were due to de Finetti, and Ramsey, who interpret
probability as a subjective quantity, personal to each individual. Whereas de Finetti interprets
probability as a two-sided bet assuming a linear utility
for money, Savage, motivated by Ramsey takes an axiomatic approach.

Savage's approach to personal probability was modeled after von-Neumann
and Morgernstern's axiomatic development of utility theory. This approach, is the most widely
referenced approach to personal probability; it has as its
foundation, behavioristic axioms of choice. Perhaps it is not too well recognized that these axioms
appear to be rooted in Bernstein's qualitative comparison
of events; save for the feature that they pertain to choices between actions in the face of
uncertainty.

A striking feature of Savage's
axioms is that their consequences lead to the simultaneous existence of both, a subjective
probability and a utility, and the maximization of expected utility
as a recipe for decision making under uncertainty. Savage's subjective probability conforms to
the Kolmogorov axioms; however, in the latter's set up, conditional
probability is a definition, whereas in the former, it is a consequence of the Savage axioms.

\section*{ Acknowledgements
}
The work reported here was supported by a grant from the City University of Hong
Kong, Project Number 9380068, and by the Research Grants Council Theme-Based Research Scheme Grant
T32-102/14N and T32-101/15R. Comments by Glen Shafer and Fabrizio Ruggeri have helped correct some inaccuracy in the orginal versions.

\end{document}